# Midlatitude ionospheric F2-layer response to eruptive solar events-caused geomagnetic disturbances over Hungary during the maximum of the solar cycle 24: a case study

K. A. Berényi [*,1,2], V. Barta [2], Á. Kis [2]

[1] *Eötvös Loránd University, Budapest, Hungary*

[2] *Research Centre for Astronomy and Earth Sciences, GGI, Hungarian Academy of Sciences, Sopron, Hungary*

# Abstract

In our study we analyze and compare the response and behavior of the ionospheric F2 and of the sporadic E-layer during three strong (i.e., Dst <-100nT) individual geomagnetic storms from years 2012, 2013 and 2015, winter time period. The data was provided by the state-of the art digital ionosonde of the Széchenyi István Geophysical Observatory located at midlatitude, Nagycenk, Hungary (IAGA code: NCK, geomagnetic lat.: 46,17° geomagnetic long.: 98,85°). The local time of the sudden commencement (SC) was used to characterize the type of the ionospheric storm (after Mendillo and Narvaez, 2010). This way two regular positive phase (RPP) ionospheric storms and one no-positive phase (NPP) storm have been analyzed. In all three cases a significant increase in electron density of the foF2 layer can be observed at dawn/early morning (around 6:00 UT, 07:00 LT). Also we can observe the fade-out of the ionospheric layers at night during the geomagnetically disturbed time periods. Our results suggest that the fade-out effect is not connected to the occurrence of the sporadic E-layers.

*Corresponding author. Tel.: +3670 3198192.
E-mail adresses: berenyik@ggki.hu (K. Berényi), bartav@ggki.hu (V. Barta) akis@ggki.hu (Á. Kis).

# 1. Introduction

Perturbations and changes in the ionosphere are caused by multiple effects. The most significant and well documented effects are those that occur during periods of global geomagnetic storms (Mendillo and Narvaez, 2010). Geomagnetic storms are time periods that are characterized by significantly enhanced geomagnetic activity that are driven by disturbances in the solar wind. Previous studies found two kinds of events which can induce geomagnetic storms (e.g. (Denton et al., 2009; Gonzalez et al., 1994; Tsurutani et al., 2006): 1. Coronal Mass Ejections (CME) and 2. Corotating Interaction Region (CIR)/ High-Speed Solar wind streams (HSS).  During the maximum of a solar cycle the CME-related geomagnetic storms are lot more common. The most intense geomagnetic storm of the solar cycle 24 maximum (examined in this study) is also connected to CME.

Several definition exist to determine a geomagnetic storm and its magnitude. Previous studies have typically identified and characterized the storm events considering the magnitude of geomagnetic indices such as *Kp* or *Dst* (Kane and Makarevich, 2010). Gonzalez *et al.* (1994) defined a geomagnetic storm as a period in which the Dst-index exceeds some key thresholds, as follows: events with a $Dst_{min}$ between −30 nT and −50 nT are *weak storms*, between −50 nT to −100 nT are *moderate storms*, and events with a $Dst_{min}$ less than −100 nT can be considered as *intense storms*. For storms with $Dst_{min} <$ -300 nT the term *super-storm* is used (Burešová and Laštovička, 2007). There is another subdivision of intense storms: the *strong* $(-200 \leq Dst_{min} \leq -100 \, nT)$, *very strong* $(-350 \leq Dst_{min} \leq -200 \, nT)$, and *great* $(Dst_{min} \leq -350 \, nT)$ storms (Danilov, 2013).

A geomagnetic storm generate a so-called ionospheric storm in the ionosphere, which has a similar evolution and phases as a geomagnetic storm, but with a faster course. The scientific community agreed, that only the large geomagnetic storms induce global ionospheric storms (Buresova et al., 2014). The general course of the midlatitude ionospheric F2-layer response to geomagnetic storms has been described by Rishbeth and Field (1997) and summarized by Prolss (2004) in the recent years. The *initial phase* starts with enhanced peak electron density, which last for few hours after the SSC of the geomagnetic storm. During the *main phase,* which lasts for a day or more, the electron density ($N_e$) is decreased compared to quiet day values (negative storm phase), but sometimes it is enhanced (positive storm phase). Lastly the *recovery phase* of the storm can last from hours to days.

The midlatitude ionosphere provides a good location to study different, comparable effects caused by multiple mechanisms. The impacts can origin from lower and higher latitudes too. Thus although the investigation of the ionosphere at midlaltitude is fraught with difficulties, it can increase our general understanding of the background physical processes and how they act paralelly.

There are several processes that have to be taken into consideration during the examination of the midlatitude ionosphere, as follows: photo-production, chemical loss, and transport by thermal expansion, neutral winds,  waves, tides and electric fields of internal and external origin (Mendillo and Narvaez, 2009). All of these come up during disturbances. Further influential factors, that need also to be taken into account during ionospheric studies (Immel and Mannucci, 2013; Mendillo and Narvaez, 2010, 2009): local time of SSC, the magnitude of the geomagnetic storm (based on *Kp*- and *Dst*-index values), the type of the geomagnetic storm (CME-related or CIR/HSS-related), the phase of the solar cycle, the geomagnetic latitude and longitude of the station and the season of the year (i.e., winter or summer).

The "smoothed" picture of the F-region changes is widely accepted since the end of the 1990s (see reviews of Prölls, 1995; Buonsanto, 1999; Mendillo, 2006), but the global distribution of ionospheric storm effects is rather complicated and differs considerably from storm to storm (Danilov, 2013). So according to the observations it can be concluded, that both a significant electron density decrease and an anomalous density increase can occur in the F2-layer over a course of a magnetic storm (Buresova et al., 2014). Also it has been observed, that there are cases when significant magnetic disturbances did not followed by ionospheric disturbance, or vice versa (Buresova et al., 2007; Field et al., 1998; Li et al., 2012).

The analysis of the sporadic E-layer (Es-layer) can be informative during intense storm event. The blanketing behavior of this layer can be in some cases responsible for the fade out of upper ionospheric layers like F1- and F2-layer. The formation of this layer is connected to the penetrated meteoric metallic and molecular ions into the E-layer heights (90-120 km), which generate a layer thanks to the wind shear of neutral waves (Pietrella and Bianchi, 2009; Whitehead, 1989). The Es-layer is formed by "patches" in the E-layer with larger electron density. The lifetime of these Es-layers are found to be between 2-10 h (see Harwood, 1961; Tanaka, 1979; Houminer *et al.*, 1996). The solar and geomagnetic influence on the midlatitude Es-layer parameters has been studied in the last decades (Baggaley, 1985, 1984; Maksyutin and Sherstyukov, 2005; Pietrella and Bianchi, 2009; Whitehead, 1989, 1970): it was shown, that positive, negative and also no correlation of Es-layer parameters can be expected in the function of geomagnetic activity. The correlation may vary with intensity of Es-layer, season and time of the day (Maksyutin and Sherstyukov, 2005). Further investigations are required to understand the behavior of this layer.

In this preliminary study we focused on the determination of the occurring ionospheric storm patterns in the F2-layer at a Hungarian ionosonde station during the most intense geomagnetic storm event of the solar cycle 24 maximum. It is important to mention that this current solar cycle is weaker than the previous ones therefore we do not have so much strong CME impacts which can generate intense geomagnetic storms. Previous studies mostly used superposed epoch analysis (SEA) or median values to examine the ionospheric variation and the magnitude of the changes caused by various effects (Kane and Makarevich, 2010; Mendillo and Narvaez, 2010). The main goal of our investigation is to compare/supplement the previous results (of e.g. Mendillo and Narvaez, 2009, 2010) that were based on statistical studies. In our study we analyze individual events in order to observe the specific patterns because the averaging tends to eliminate the individual signature of the event. Besides the verification of the previous studies and the accepted general features occurring during intense ionospheric storm events are important and necessary when we would like to develop the accuracy of the space weather prediction models. The unique/specific features that appear during an individual storm can be important and can give us new information about the processes.

## 2. Analysis methods

The data provided by the state-of the art ionosonde located at the Széchenyi István Geophysical Observatory in North-West Hungary, in Nagycenk (IAGA code: NCK) was used for the analysis. Its McIllwain number is L =1.9 therefore this station is situated at optimal (magnetic) latitude to observe the ionospheric processes in a midlatitude region. The other important information about the station: geomagnetic latitude: 46,17°, geomagnetic longitude: 98,85°, inclination (dip angle): 66,83°. The ionosonde at the NCK observatory is a VISRC-2 type ionosonde (for a detailed description of the ionosonde see the article by Sátori *et al.*, 2013*).*
The relevant ionospheric parameters were determined after a careful manual evaluation of the ionograms. During the investigation we focused on the foF2 parameter because this parameter indicates the maximum plasma (electron) density in the F2-layer. We also examined the foEs parameter (maximum electron density of sporadic E-layer) and h'F2 parameter (virtual height of the F2-layer) in order to study the ionospheric processes during geomagnetic storms. The studied time interval covers three years of maximum phase of the solar cycle 24, i.e., from 2012 to 2015, in winter time period. The biggest and most effective CME events occure in this solar cycle phase, during minimum phase the occurrence and magnitude of the CME events decrease and mostly HSS/CIR events can be observed. The winter ionosphere is different from the summer one, the background (solar-induced) thermospheric wind circulation is opposite to the storm-induced in the sunlit hemisphere, and therefore generally positive ionospheric storm phases can be expected at midlatitude. However, at night the circulations coincide, and rather frequent occurrence of negative ionospheric storm phase are registered in the main phase of the geomagnetic storm (Buonsanto, 1999; Danilov, 2013).

We also used solar and geomagnetic indices (hourly data) provided by OMNIWEB.

In order to compare the geomagnetically quiet days with disturbed (stormy) days, we needed reference values to characterize a geomagnetically quiet day. Therefore, 5 geomagnetically quiet (Kp-index < 2) days have been chosen for reference (in the 2013/01/01-2013/01/05 time period) and the half hour averages have been calculated using the data of the five quiet days to obtain an average quiet day ionospheric parameter change vs time (**Figure 1**).

The three most intense CME-related geomagnetic storm events have been chosen as geomagnetically disturbed time periods for the analysis. These events were observed during the following time periods: 2012/11/11-17 ($Dst_{min}$ = -108 nT), 2013/03/16-23 ($Dst_{min}$= -132 nT) and 2015/03/16-25 ($Dst_{min}$= -228 nT). To cover completely the disturbed time period the selected intervals contain the pre-storm (24 h before the SSC), the initial, the main storm and the recovery phase of each event. To identify the storm phases, we used the Dst-index. The events were followed until the Dst-index returned above -10 nT. The SSC times of these storms is obtained from the official dataset of SSCs ([ftp://ftp.gfz-potsdam.de/pub/home/obs/kp-ap/ssc.dat](ftp://ftp.gfz-potsdam.de/pub/home/obs/kp-ap/ssc.dat)). All of these storms started during the night in wintertime, therefore the comparison of these events is relevant, because the physical processes generated by a geomagnetic storm strictly depend on the LT of the SSC (Mendillo and Narvaez, 2010, 2009).

## 3.1. Observations and results

**The geomagnetic storm from 2012, November 11-17.**
The magnitude of this storm was (on 14$^{th}$ of November.): $Dst_{min}$ = -108 nT, $Kp_{max}$ = 6,25. The Sudden Storm Commencement (SSC) was at 23:12 UT (00:12 LT) on day 12 of the month. This is a somewhat unusual storm because the main phase of the storm reached its maximum only one day after the arriving of the SSC, as it can be clearly seen on **Figure 2,** which makes difficult to categorize its type using the rule of Mendillo and Narvaez, 2009.

On the other hand, some characteristics during the storm can help us in the categorization: during the main phase of the storm a negative ionospheric (storm) phase can be observed, which is followed by an increase in the foF2 parameter during the early recovery phase of the storm, so this storm presents the characteristics of a No-Positive Phase (NPP) storm. Note: during this storm the ionosonde in NCK provides data only up to 8 MHz, but the negative ionospheric response during the main phase of the geomagnetic storm can be observed clearly.

The negative response in the foF2 parameter during the main phase of the geomagnetic storm is a known response of the ionosphere in winter, but usually (not only) during stronger storms. In this case we can observe a negative ionospheric storm phase during a relatively weak (i.e., closer to moderate) geomagnetic storm, this event can be considered as an atypical storm, during which the main phase and also the ionospheric storm's phase is delayed. In a regular storm, the main phase should start with a significant positive or negative phase within a few hours after the SSC.

This effect is not so common in winter at midlatitude. Kane (2005) also studied a similar storm (he called it event Z intense geomagnetic storm 1989/03/13-14) and he found a similar magnitude of negative ionospheric response in the main phase of the geomagnetic storm in every foF2 data of surrounding ionosondes. Also he registered pre-storm enhancement, 24 h before the SSC time (like Burešová and Laštovička, 2007), but in our storm case the magnitude of this effect does not seem significant. Because of the missing data this conclusion could be strengthen with data of neighboring ionosonde stations (e.g. Pruhonice). On the early recovery phase the slow recovery of the electron density can be seen, so the early recovery phase enhancement effect just slightly can be seen (see **Figure 2**). In the main phase (on 2012/11/14) at night from 20:00 to 00:30 UT, disappearance of the ionospheric layers can be detected on the ionograms (**Figure 2**) which indicates that the electron density of the F-layer decreased below the detectability level. The ionosondes detect the signals from 1,5 MHz. Our results might suggest that the effect is more common at midlatitude by negative ionospheric phases at night during the main phase of the geomagnetic storms. Besides the magnitude of the storm might be also an important factor, supposedly only intense geomagnetic storms ($Dst_{min} < -100$ nT) can induce such effects.

The possible reason for the negative phase during the main storm phase: as a result of geomagnetic storms, the auroral heating induced composition disturbance zone (with decreased $O/N_2$ ratio, which cause the electron density depletion) propagate in the ionosphere from the auroral region to lower latitudes (Buonsanto, 1999). The dynamical regime of the thermosphere, which affect the propagation of this composition disturbance zone, is different in summer and winter. In winter in the sunlit sector the background thermospheric wind circulation (solar-induced, directed poleward) is opposite to storm-induced circulation (directed equatorward) (Danilov, 2013). Therefore, there is a latitude zone, where the negative phase stops. In this 2012 storm case, the composition disturbance zone possibly reached our midlatitude station (**Figure 2**).

There is a rapid electron density decrease at night after the main phase of the storm: at night the two circulations coincide, and this leads to rather frequent occurrence of the negative phase at midlatitudes (Danilov, 2013). In **Figure 2** this negative phase can be clearly seen during the night time hours. The dynamical change in the thermospheric circulation should not cause such strong decrease (the F2-layer disappear from the ionograms, see below in **Figure A1**). Another process might contribute to this strong decrease in the electron density at night: during storms our station's L=2 domain often remain within the plasmasphere during daytime hours (contributing to daytime positive phase ionospheric storms). On the other hand, after sunset our midlatitude region is beyond the plasmapause, mostly during strong storm events in the day of the main phase (Mendillo and Narvaez, 2010). This loss of plasmaspheric connection, which most of the time streghten the positive phase ionospheric storms during sunlit hours, lets the chemical processes (decreased $O/N_2$ ratio) to be more effective and evoke strong negative ionospheric storm phase at night (see Mendillo, Klobuchar and Hajeb-hosseinieh, 1974; Mendillo and Narvaez, 2009).

### 3.1.1. Comparison of storm 2013 and 2015 (St. Patrick's Day events)

**The geomagnetic storm from 2013, March 16-23.**
The magnitude of this storm was (on $17^{th}$ of March): $Dst_{min}$ = -132 nT, $Kp_{max}$ = 6,7. This geomagnetic storm's SSC time was at 06:00 UT (07:00 LT) on $17^{th}$ of March.
In the main phase of the storm the electron density increase is significant (**Figure 3**). During the early recovery phase a moderate increase in the electron density value can be seen, while on the last days of the time period the difference between the mean quiet day curve and the observed curve becomes negligible. This case is a "school book" example for a regular positive phase (RPP) ionospheric storm type according to the description by Mendillo and Narvaez, 2010. The local time of the SSC was at 07:00 and immediately after that a strong positive phase is registered.

**The geomagnetic storm from 2015, March 16-25.**
The magnitude of this storm was (on day $17^{th}$ of March.): $Dst_{min}$ = -223 nT, $Kp_{max}$ = 7,67. More CME information in the article of Wu *et al.* (2016). This geomagnetic storm's SSC was at 04:45 UT (05:45 LT) on day $17^{th}$. The storm, also known as St. Patrick's Day storm, is the largest storm of the solar cycle 24. Just like the storm from 2013, this event also generated a Regular Positive Phase (RPP) ionospheric storm. The patterns are also quite similar in the case of these two events (see **Figure 4**).

The storms from 2013 and 2015 by coincidence start at the same date, only the time of the SSC differs with 1 hour and 55 minutes if we compare the two cases. On day $16^{th}$ a significant pre-storm enhancement (Burešová and Laštovička, 2007) can be seen in the foF2 parameter in both storm cases. The local time of the SSCs were at 07:00 and at 05:45 (these are marked with red dotted line on the **Figure 3** and **4**). Immediately after the SSC the positive phase of the ionospheric storm started. Depletions in the foF2 parameter value compared to the main phase value can be observed on the next days. These storms are as Regular Positive Phase (RPP) storms because in the main phase of the storm, electron density increase can be seen, with no delay (Mendillo and Narvaez, 2010). In wintertime this is the generally observed type of the ionospheric storm at midlatitude (Buonsanto, 1999; Danilov, 2013). As the article of Danilov (2013) concluded, negative phases occur in all season except winter, while positive phases are more probable at winter. The latitude-connected physical processes (see the article of Buonsanto, 1999; Danilov, 2013) cause that almost negative phases can be observed at high latitudes and positive phases tend to occur at middle and low latitudes.

An increase of the foF2 parameter, which is called pre-storm enhancement in the literature (Burešová and Laštovička, 2007; Danilov, 2013; Kane, 2005) was observed the day before the SSC in both cases. In the article by Burešová and Laštovička (2007) the authors reach to the conclusion that the origin of these pre-storm enhancements can't be determined. The pre-storm enhancement feature can be possibly used as storm precursor (24 h before the SSC), as Kane (2005) and Blagoveshchensky, MacDougall and Piatkova (2006) discussed. According to the paper of (Burešová and Laštovička, 2007) the following processes can possibly cause this effect: solar flares (they can only occasionally strengthen the pre-storm enhancements), soft particle precipitation in the dayside cusp, magnetospheric electric field penetration, auroral region activity expressed via the AE index, and Mikhailov's quiet-time F2-layer disturbances. The magnitude of the change registered in the foF2 value does not seem to be directly correlated with the magnitude of the geomagnetic storm (see the Dst-index on **Figure 3** and **4**). In other words, a significantly larger Dst value does not result in a significantly different response in the foF2 value as it can be seen in **Figure 5**. During both storms the foF2 parameter reaches the maximum value of ~12,5 MHz after the SSC around the same time (at 11:30 UT) and later in both cases a pronounced "dusk effect" can be observed during the afternoon hours (**Figure 5**). On the other hand, during the main phase of the storm a quite significant difference can be seen in the nighttime period of the two events. The foF2 parameter decreases much faster in the evening hours during the storm from 2015 and it even disappears at 24:00 UT on 17$^{th}$ of March (**Figure 4**). This disappearance of the foF2 layer occurs in the following two nights also during the 2015 storm, while a similar effect cannot be observed during the 2013 storm. Note: a similar effect can be observed during the storm from 2012, i.e., the fade-out of the ionospheric layer. However, during the 2012 storm this effect is much more significant and has a longer duration. Most probably this is due to the fact that the 2012 storm was a negative ionospheric storm.

During the recovery phase of the 2015 storm (from 21/03/2015 to 23/03/2015) a significant increase can be observed in the foF2 value, which is unique among the three cases. This electron density increase can be explained by a subsequent geomagnetic disturbance (see **Figure 5**). This second storm does not appear in the value of the Dst-index, but it is clearly visible in the Kp value. This disturbance, which occurred during the recovery phase of the first storm, generated a positive ionospheric storm with a pre-storm effect too. The "dusk effect" also can be observed after the (second) SSC in the afternoon sector. The positive phase -with a significantly elevated foF2 value- persisted for a long period of time: it can be observed even four days after the (second) SSC (see **Figure 4**), which makes this storm a very unusual and unique one.

### 3.1.2. ΔfoF2, h'F2, foEs parameter examination during the three storm events

In order to study in detail the differences between the geomagnetically disturbed and the quiet days, we performed a quantitative analysis. We calculated the residuals in percentage using the following formula (after Buresova *et al.*, 2014):

$$\Delta foF2 = \left(\frac{foF2_{storm} - foF2_{quiet}}{foF2_{quiet}}\right) * 100\ \% \qquad (1)$$

With this equation the difference between the mean quiet day value and the storm time value can be obtained. When the value on **Figure 6** is 0 %, the storm time value is equal to the quiet day value at the respective hour.

The deviations of the storm time values from its quiet day average pair shows us the magnitude of the effects during geomagnetic storms. The maximum peak of the electron density increase, which appears at 11:30 UT (12:30 LT) on **Figure 3** and **4,** is smaller on **Figure 6 (b)** and **(c)**. Because the magnitude of the deviation from the quiet day curve at this time is not as large as during the afternoon hours.

On **Figure 6 (b)** and **(c)** the electron density increase on the day after the SSC in the afternoon hours is the most remarkable phenomenon , which is called "dusk effect" in the literature (Buonsanto, 1999; Kane, 2005). The magnitude of the "dusk effect" seems to show a proportional connection with the magnitude of the geomagnetic storms. It is striking that this proportionality is valid in all three cases, irrespective of the type (i.e. positive or negative) of the storm. The values can be two times larger than at its quiet day pair like around 17:00 UT (18:00 LT) on **Figure 6** (**c**). However, we have to mention that the validation of this result need further investigation, which will be made with a statistically appropriate amount of storm cases. Furthermore, in this two cases a large peak of electron density increase appears around 6:00 UT (7:00 LT) in both cases approximately with the same magnitude at the same time with the SSC too (**Figure 6 (b)** and **(c)**). This effect appears also in the data of storm 2012, but with a much less magnitude (see **Figure 6 (a)**). In all three cases this can be interpreted as a kind of dawn effect.

It is important to mention that the magnitude of the changes are larger compared to the values presented by Mendillo and Narvaez (2009, 2010) who performed statistical analysis of multiple events. The difference shows the importance of the detailed studies that focus on individual events because the results suggest that a statistical analysis tend to remove the highly variable characteristics of an ionospheric response to a geomagnetic disturbance.

h'F2 parameter:

The **Figure 7** presents the h'F2 parameter fluctuation vs time for the three selected storm events including the pre-storm, the main phase and the early recovery phase of the geomagnetic storm. During the 2012 storm event one can observe a significant increase of the h'F2 parameter in the main phase, during the sunlit period. This support our assumption that in this case the plasma of the F2-layer moved along the magnetic field line, generated by the auroral heating caused shock wind, which caused the uplifting of the layer (Danilov, 2013). Besides the shock wind generated thermospheric wind circulation (storm-induced) transports the composition disturbance zone with decreased $O/N_2$ ratio (that enhances loss) to lower latitudes. In winter, this process rarely reach the midlatitude region, because of the aforementioned fact, that the background thermospheric wind circulation opposite to the storm-time circulation during daytime hours, and from this reason the equatorward propagated composition disturbance zone stops in most cases at higher latitudes. However, strong geomagnetic storms can move this disturbance zone to lower latitudes.

At night, there is a change in the direction of the background circulation (see above). Therefore, the uplifted F2-layer plasma with the composition disturbance zone can propagate much lower latitudes. This effect nicely can be seen in all three storm cases in both foF2 and h'F2 parameter (see **Figure 2, 4 and 7**): the foF2 parameter decreases, the h'F2 inreases. The increased erodation of electron density in the F2-layer, can cause the fade-out of the layer at night. On **Figure 7** this erodation can be seen also in h'F2 parameter (not just in foF2, see **Figure 2** and **4**) by storm 2012 and 2015. The F2-layers disappear from the ionograms, because its intensity go under 1,5 MHz, which is the detectability level of the ionosonde.

One have to be careful with the analysis of the h'F2 parameter (virtual height of the F2-layer), because it is widely known that it is not a reliable indicator of the real F2-layer heights. Therefore the conclusions based on this parameter can be erroneous, but still the tendency/movement of the F2-layer height can be followed with this parameter too.

foEs parameter:

**Figure 8** presents the evolution of the foEs and the foF2 parameter versus time during calm days and during the pre-storm and main phases of the analyzed disturbed time periods. The first obvious conclusion to be drawn from the evolution of the foEs and the foF2 parameter is that the presence of the Es-layer does not influence the upper F2-layer above it. In other words the Es-layer does not mask the F2-layer, therefore the observed fade-out effect is not connected to the occurrence of the sporadic E-layer. On the other hand, the magnitude of the geomagnetic disturbance do has a detectable effect on the occurrence of the sporadic E-layer: with increasing magnitude the occurrence of the Es-layer less. One can observe that during the largest storm (i.e., from 2015, **Figure 8**, lower plot) the Es-layer disappears almost completely. Consequently, these results support the occurrence of the negative effect in the function of geomagnetic activity. On the other hand, Pietrella and Bianchi (2009) concluded that this effect is not significant, while the result of our study shows the opposite.

Maksyutin and Sherstyukov (2005) analyzed the geomagnetic influence on the midlatitude Es-layer and found that the decrease in the layer intensity can be observed in summer and winter during the day of the maximum geomagnetic disturbance. Our results suggest that a certain magnitude of the geomagnetic disturbance is needed to decrease the Es-layer's intensity below the detectability level of the VISRC-2 ionosonde, and in addition, the disappearance of the Es is recorded mostly during the night (see **Figure 8**) during intense geomagnetic storms. In the case of the stronger disturbances (e.g. 2015 geomagnetic storm), the disappearance of the Es-layer becomes visible even during the daytime and persist for several days and not only on the day of the maximum disturbance.

Further investigation is needed to validate our conclusions about foEs parameter.

# Conclusions

In our study we present a case study of the response of the mid-latitude ionospheric layers to geomagnetic storms with different magnitudes by analyzing in detail three different geomagnetic storms from the maximum period of the solar cycle 24.

Several known phenomena can be seen in the variation of the foF2 parameter of the three selected geomagnetic storm event: the **pre-storm enhancement**, the **"dawn effect"** at around 06:00 UT (07:00 LT) and the **"dusk effect"** in the afternoon at around 18:00 UT (19:00 LT) hours and the **early recovery phase enhancement**.

In order to quantify the effect of the geomagnetic disturbance, the deviation of the foF2 from an averaged quiet day value has been determined for each of the three cases.

At dawn there is a positive effect, which does not seem to be affected by the magnitude of the geomagnetic storm (characterized by the Dst-index) while the dusk effect shows a clear proportional relation with strength of the geomagnetic storm. It is important to mention that the deviation from the quiet day main value for these individual events is significantly more than the previous deviation values that are based on statistical analysis.

The fade-out of the ionospheric layers were detected in the main phase of the analyzed winter time geomagnetic storms in the following cases: in the storm 2012 when the ionospheric storm phase was negative and the $Dst_{min} < -100$ nT, in the storm 2015, when a positive ionospheric storm was generated and the $Dst_{min} < -200$ nT. The analysis suggests that the effect is more common at midlatitude during negative ionospheric phases at night at the main phase of the geomagnetic storm. Furthermore our results confirm the commonly accepted idea is that only intense geomagnetic storms ($Dst_{min} < -100$ nT) can induce such effects. The h'F2 and foEs parameter analysis support this statement. The h'F2 parameter value increase significantly before the fade-out effect, therefore a fast uplifting of the F2-layer can be observed at nighttime hours continuing with disappearance of the data.

Our results also show that the Es-layer cannot cause the fade-out effect because it does not blanket the F2-layer.

On the other hand, our results suggest that the magnitude of the geomagnetic storm do has an observable influence on the occurrence of the Es-layer: with increasing magnitude the intensity of the Es-layer less and it goes under the detectability level during the most intense geomagnetic storm.

# Acknowledgments

We thank the data centers (OMNIWEB and WDC for Geomagnetism, Kyoto) and the Széchenyi István Geophysical Observatory at Nagycenk, Hungary for supplying high quality data for the research. The study was partially funded by the NKFIH NN 116446 and by the HAS-JSPS NKM-95/2016 research project. Also I thank you for the reviewers for the valuable advices, which improved my article.

**CAPTIONS (with figures)**:

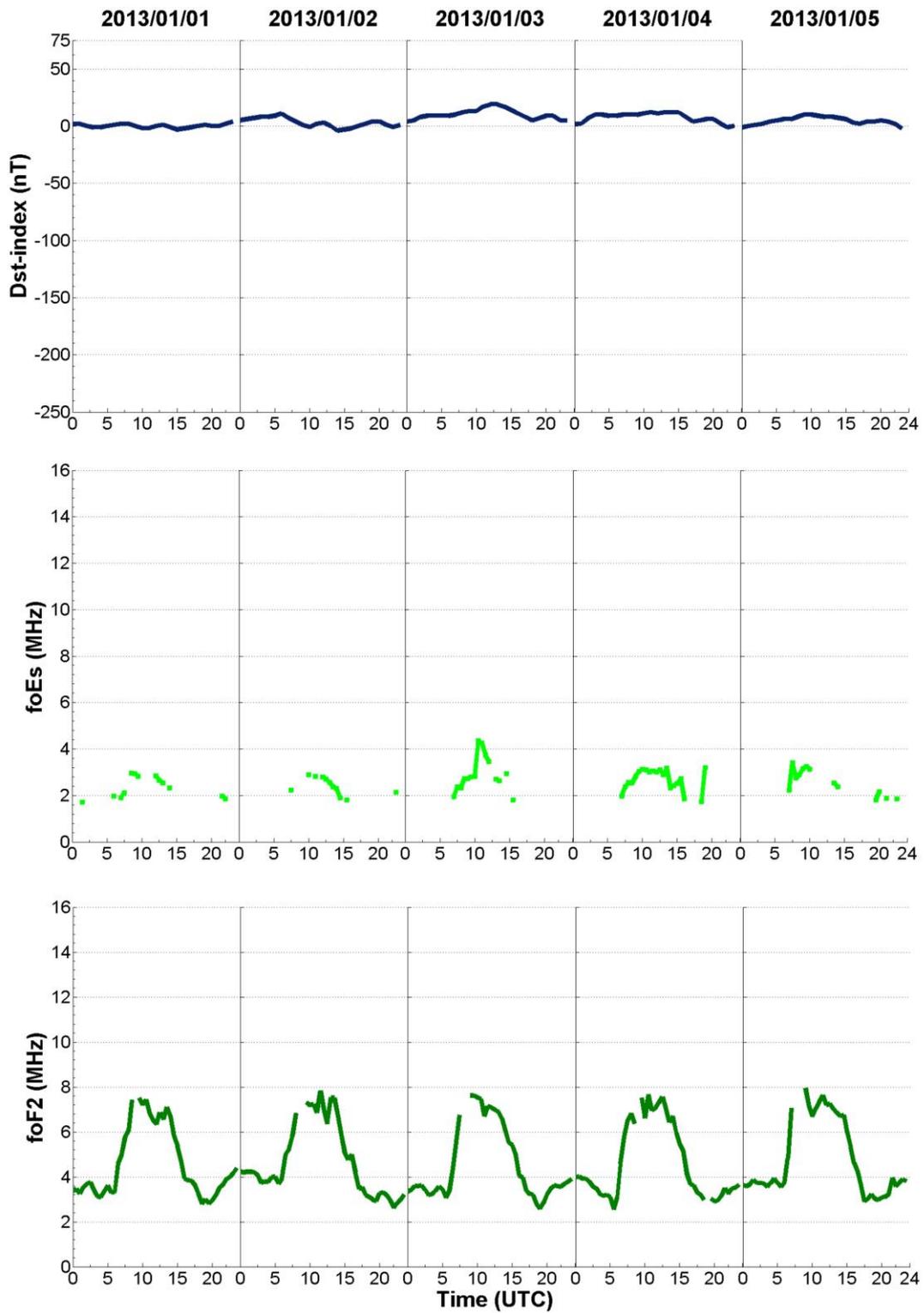

**Figure 1** caption: On the upper diagrams the Dst-index values and on the lower diagrams the foEs and the foF2 values of selected 5 calm day's are portrayed, which were used for references.

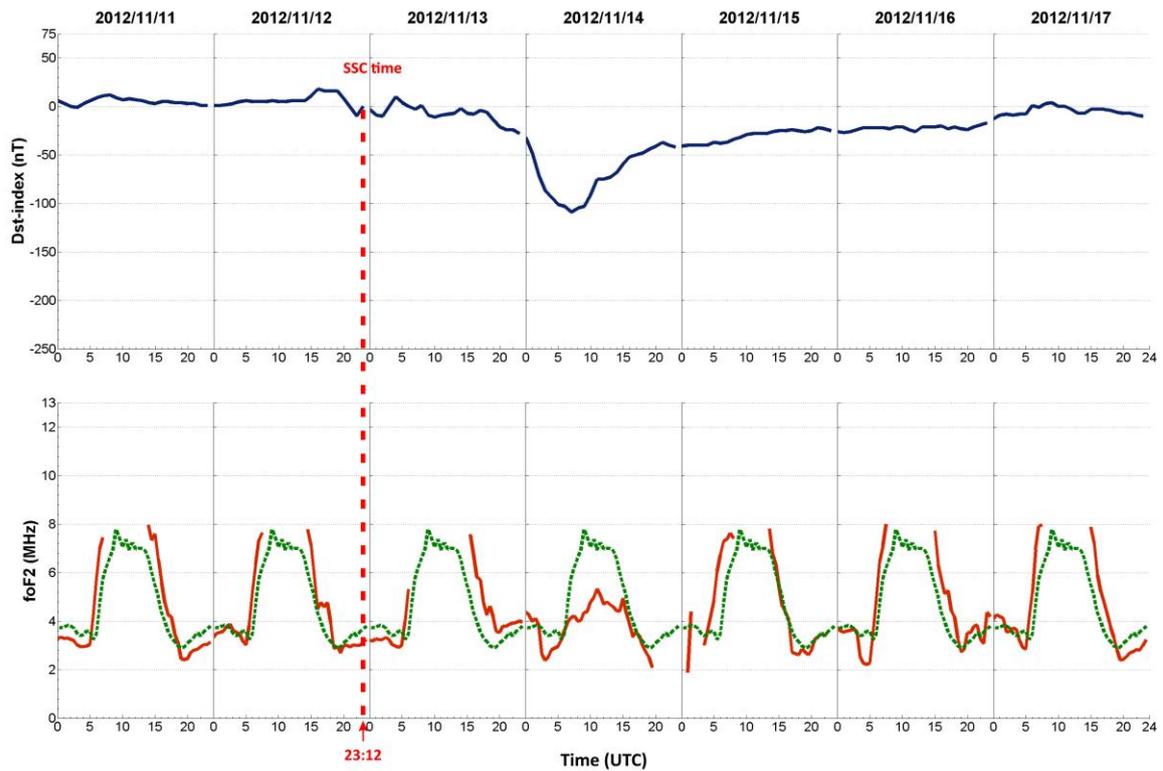

**Figure 2** caption: A full geomagnetic storm period from 2012/11/11 to 2012/11/17. The daily variation of the Dst-index (upper plot), the daily variation of the ionospheric foF2 parameter (associated with the electron density, lower plot) are portrayed. Data gaps can be observed around noon every day because the ionosonde in NCK provided data up to 8 MHz during this storm. The local time of the SSC was at 23:12 (red dotted line). The daily variation of the mean foF2 parameter generated from the reference day's data is marked with the green dotted line.

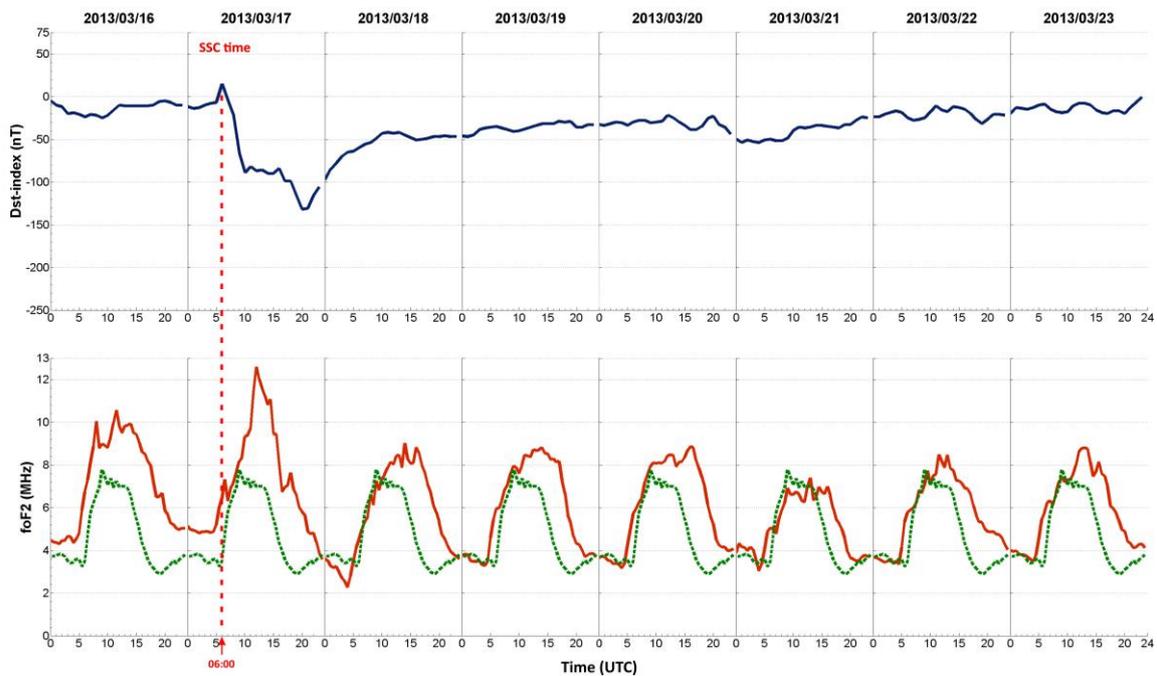

**Figure 3** caption: The full geomagnetic storm period from 2013/03/16 to 2013/03/23. The daily variation of the Dst-index (upper plot), the daily variation of the ionospheric foF2 parameters (associated with the electron density, lower plot) are portrayed. The local time of the SSC was at 07:00 (red dotted line). The daily variation of the mean foF2 parameter generated from the reference day's data is marked with the green dotted line.

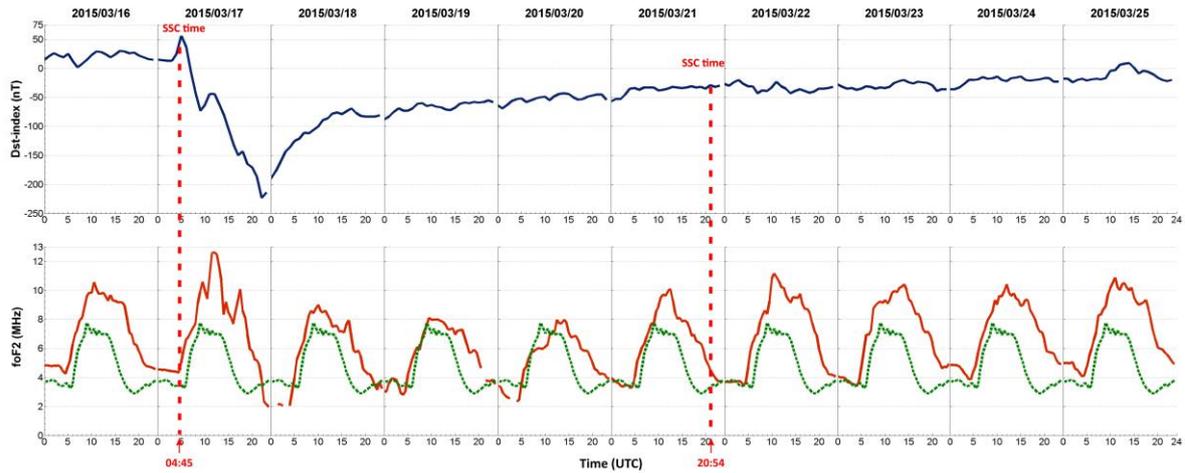

**Figure 4** caption: The full geomagnetic storm period from 2015/03/16 to 2015/03/25, the so-called St. Patrick Day's storm. The daily variation of the Dst-index (upper plot), the daily variation of the ionospheric foF2 parameter (associated with the electron density, lower plot) are portrayed. The local time of the first SSC was at 05:45 and the second SSC was at 21:54 (red dotted lines. The daily variation of the mean foF2 parameter generated from the reference day's data is marked with the green dotted line.

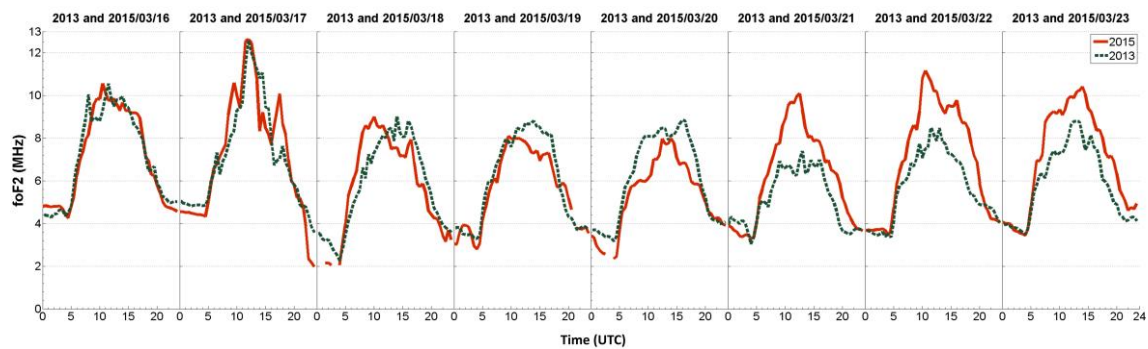

**Figure 5** caption: The foF2 values of two similar storms from year 2013 and 2015 are portrayed to compare the two cases. In the main phase significant difference can be seen at night. The magnitude of the maximal peaks of the electron density increase (positive phase) are equal in both storms. In the case of storm 2015 an increase can be seen between 21/03/2015 23/03/2015 which can be explained by a second storm (detailed in the text).

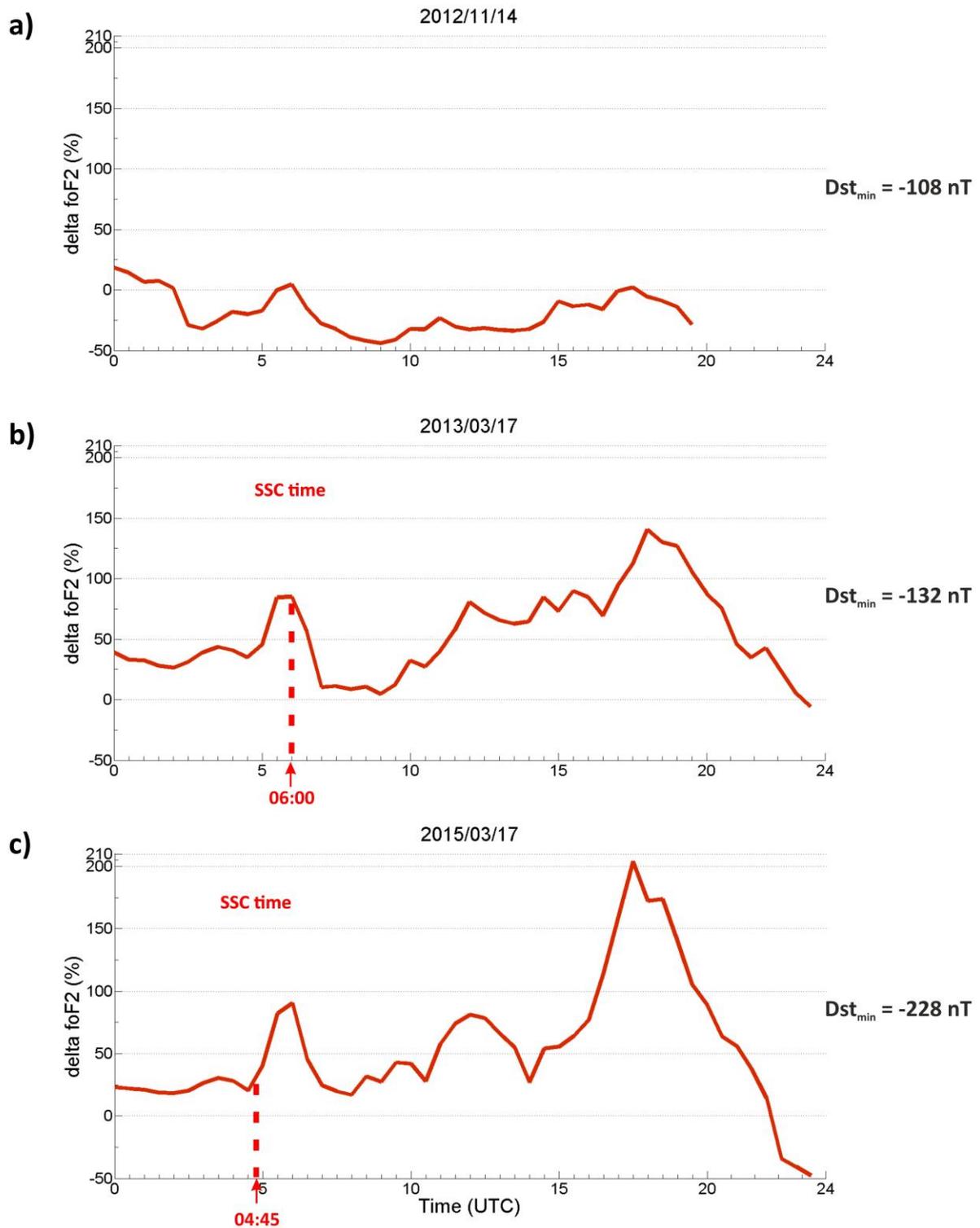

**Figure 6** caption: The percentage difference (residuals, detailed in the text) from the quiet day values are portrayed during the three storms. Only the main phase of the storms were analyzed with this method. The storm of 2012/11/14 can be seen with the negative ionospheric storm phase (a). The storm from year 2013 is portrayed (b) and the positive ionospheric storm effect also can be seen with this method. The strongest analyzed storm is portrayed from year 2015 with a significant positive ionospheric storm phase (c). The magnitude of the changes correlated with the magnitude of the storm (as the Dst-index show us). With the dotted line the time (UTC) of the SSC is portrayed on the diagrams.

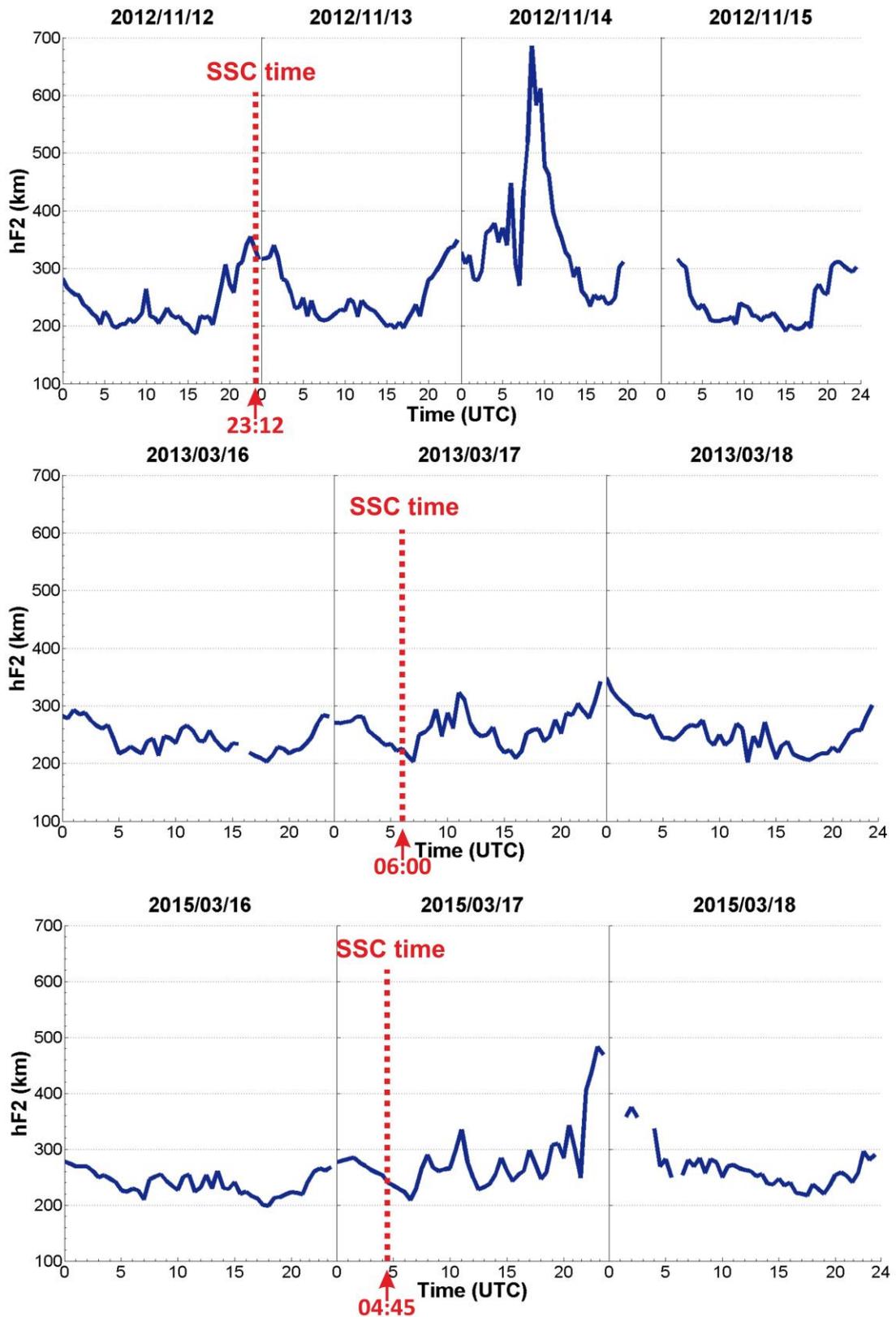

**Figure 7** caption: The h'F2 parameter variation during the three selected storm's pre-storm and main phase is plotted. The SSC time is displayed with red dashed lines.

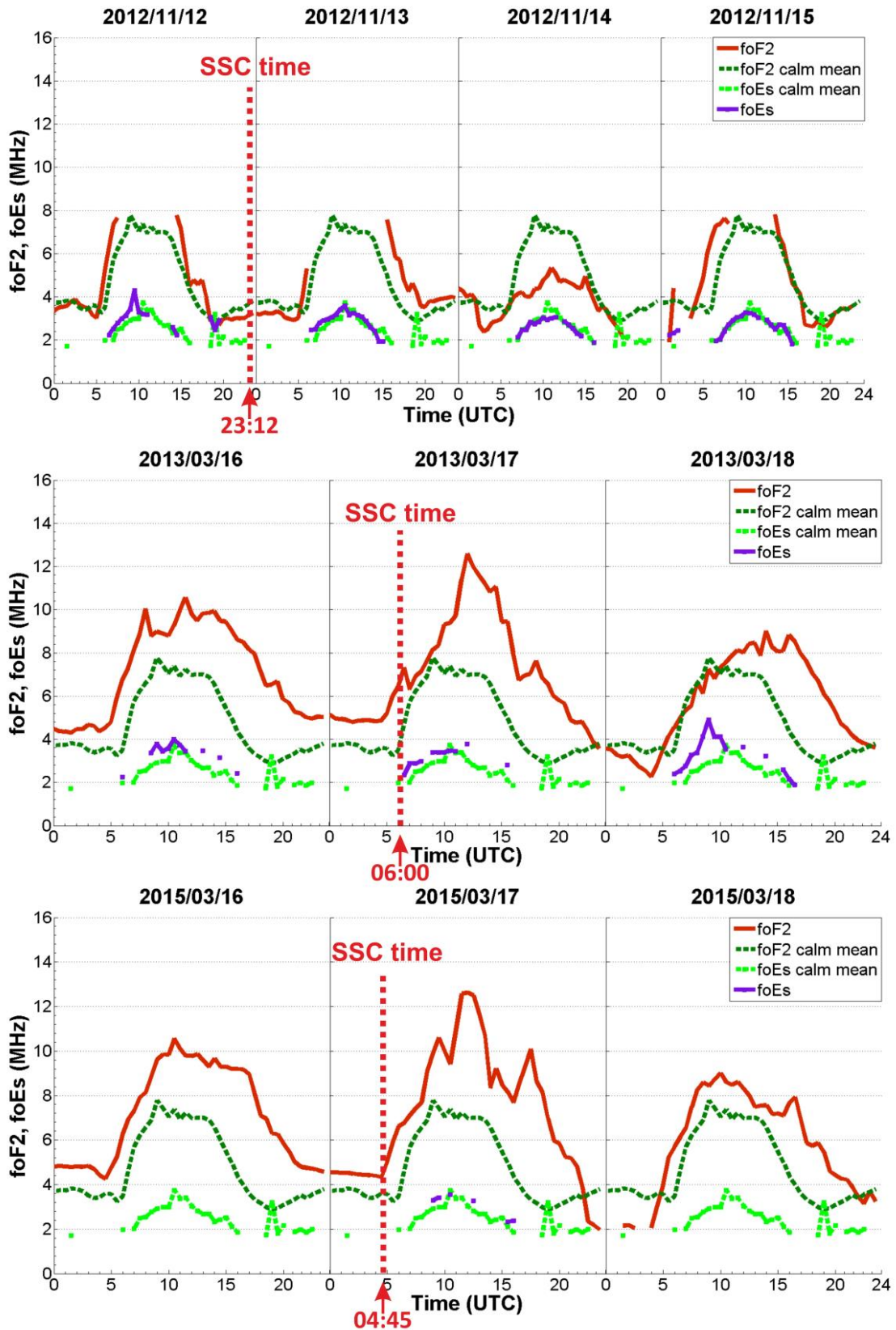

**Figure 8** caption: The foEs parameter variation (with its calm mean curve) of the three selected storm's pre-storm and main phase is plotted. The foF2 parameter fluctuation also can be seen on the figure in order to see the processes better. The SSC time is displayed with red dashed lines.

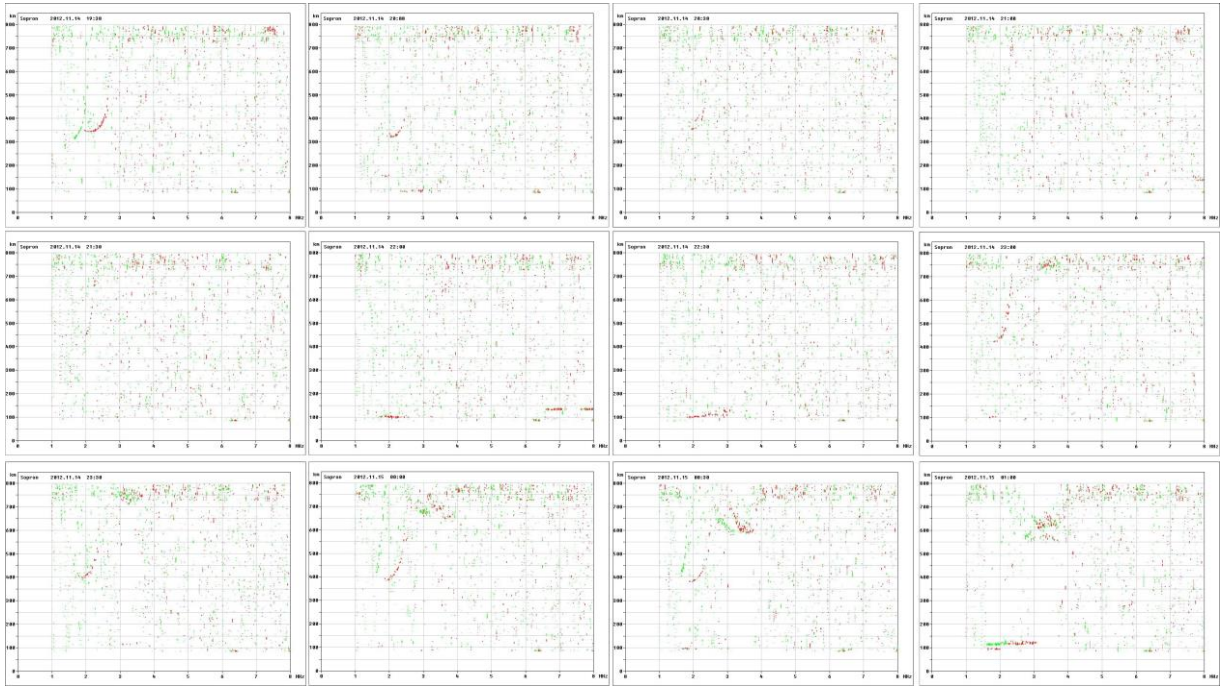

**Figure A1** caption: The ionogram sequence of the 2012/11/14 main phase from 19:30 to 01:00 UT. The fade-out of the ionospheric layers was from 20:00 to 00:30 UT. The ordinary wave mode (the scaled signal) here is green.

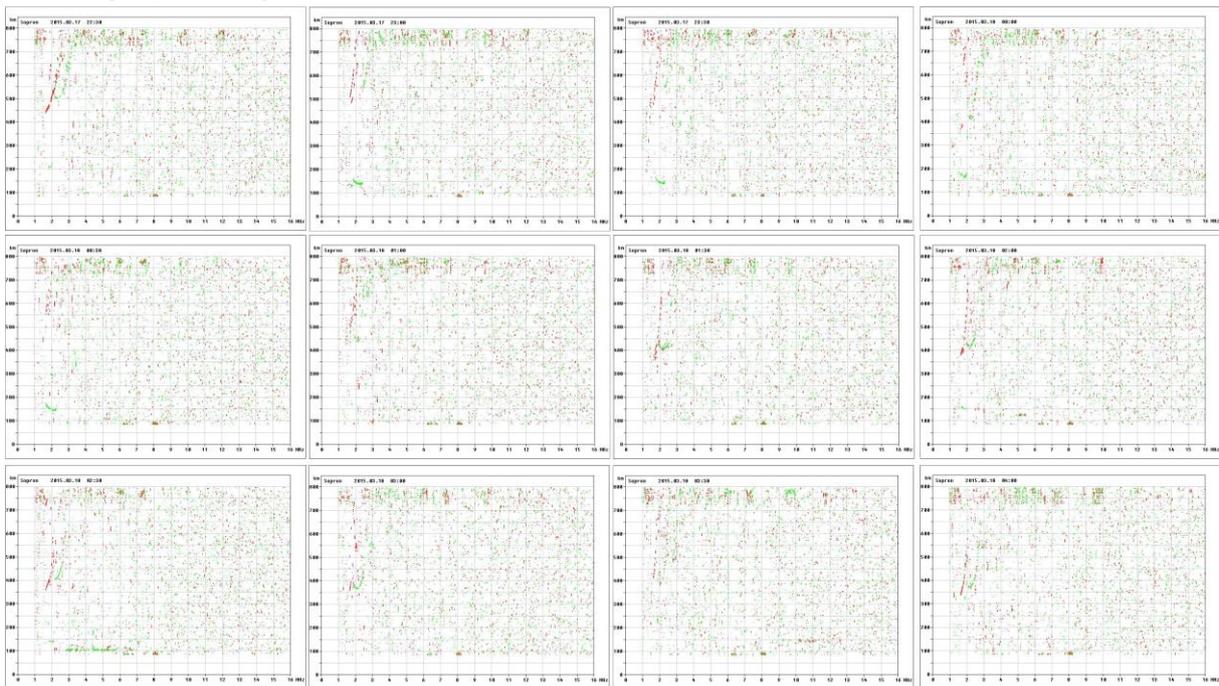

**Figure A2** caption: The ionogram sequence of the 2015/03/17 main phase from 22:30 to 04:00 UT. The fade-out of the ionospheric layers was from 0:00 to 01:00 UT. The ordinary wave mode (the scaled signal) here is red.